\documentstyle[11pt,epsfig,amssymb,amsmath]{article}
\title{\bf Quark-hadron phase transition in Brans-Dicke brane gravity }
\author{K. Atazadeh$^1$\thanks{email: k-atazadeh@sbu.ac.ir}
,\, A. M. Ghezelbash$^2$\thanks{masoud.ghezelbash@usask.ca} \,\
and H. R. Sepangi$^1$\thanks{email: hr-sepangi@sbu.ac.ir}
\\ {\small $^1$Department of Physics, Shahid Beheshti University G. C., Evin,
Tehran 19839, Iran}\\ {\small $^2$Department of Physics and
Engineering Physics, University of Saskatchewan,}\\{\small
Saskatoon, Saskatchewan S7N 5E2, Canada}}
\begin{document}
\maketitle
\begin{abstract}
A standard picture in cosmology has been emerging over the past
decade in which a phase transition, associated with chiral symmetry
breaking after the electroweak transition, has occurred at
approximately $10^{-6}$ seconds after the Big Bang to convert a
plasma of free quarks and gluons into hadrons. In this paper, we
consider the quark–-hadron phase transition in a Brans-Dicke brane
world scenario within an effective model of QCD. We study the
evolution of the physical quantities relevant to quantitative
description of the early universe, namely, the energy density,
temperature and the scale factor before, during, and after the phase
transition. We show that for different values of the Brans-Dicke
coupling, $\omega$, phase transition occurs and results in
decreasing the effective temperature of the quark--gluon plasma and
of the hadronic fluid. We then move on to consider the quark--hadron
transition in the smooth crossover regime at high and low
temperatures and show that such a transition occurs and results in
decreasing the effective temperature of the quark-gluon plasma
during the process of quark--hadron phase transition.
\end{abstract}

\section{Introduction}

Over the past decade the possibility that the observable universe is
a brane \cite{Rubakov} embedded in a higher dimensional space-time
has been the hallmark of great many research works.  This scenario
has motivated intense efforts to understand the case where the bulk
is a 5-dimensional anti de-Sitter space. In this setup, gravitons
are allowed to penetrate into the bulk but are localized on and
around the brane \cite{lisa}. It was then shown that in a background
of a non-factorizable geometry an exponential warp factor emerges
which multiplies the Poincar\'e invariant 3+1 dimensions in the
metric. The existence of branes and the requirement that matter
fields should be localized on the brane lead to a non-conventional
cosmology which has seeded a large number of studies. Of interest in
the present study are the brane-world models in the context of
Brans-Dicke (BD) gravity. Interestingly, it has been shown that in
such a scenario  and in the presence of a BD field in the bulk
\cite{mennim}  the conservation equation for the matter field can be
satisfied. It would therefore be of interest to study  the
quark-hadron phase transition in the context of BD brane world
theory. The question of quark-hadron phase transition in the context
of conventional brane-world models has been addressed previously
\cite{harko,heydari}.

Standard cosmology suggests that as the early universe expanded and
cooled, it underwent a series of symmetry-breaking phase
transitions, causing topological defects to form. It is the study of
such phase transitions that would pave the way for a better
understanding of the evolution of the early universe, characterized
by the existence of a quark-gluon plasma undergoing a phase
transition. In what follows we focus attention on possible scenarios
which might have occurred to allow such a phase transition come to
the fore. We generally follow the discussion presented in
\cite{harko} which puts the quark-gluon phase transition in a
cosmologically transparent perspective.

The existence of phase transition from the quark–-gluon plasma phase
to hadron gas phase is a definite prediction of QCD. However, the
phase transition in QCD can be characterized by a truly singular
behavior of the partition function leading to a first or second
order phase transition, and also it can be only a crossover with
rapid changes in some observables, strongly depending on the values
of the quark masses. The possibility of a phase transition in the
gas of quark–-gluon bags was demonstrated for the first time in
\cite{Gorenstein}. Most studies have shown that one may obtain
first, second and higher order transitions. In addition, the
possibility of no phase transitions was pointed out in
\cite{Greiner}. Recently, lattice QCD calculations performed for two
quark flavors  suggest that QCD makes a smooth crossover transition
at a temperature of $T_c\sim 150 $ MeV \cite{tan}. Such a phase
transition could be responsible for the formation of relic
quark--gluon objects in the early universe which may have survived.
In this paper, our study of phase transition is based on the ideas
proposed in the first reference in \cite{Gorenstein}, where it was
shown that under certain conditions a gas of extended hadrons could
produce phase transitions of the first or second order, and also a
smooth crossover transition that might be qualitatively similar to
that of lattice QCD.

The cooling down of the color deconfined quark-gluon plasma below
the critical temperature believed to be around $T_c\approx150$ MeV
makes it energetically favorable to form color confined hadrons
(mainly pions and a tiny amount of neutrons and protons, since the
net baryon number should be conserved). However, such a new phase
does not form promptly. Generally speaking, a first order phase
transition needs some supercooling to achieve the energy used in
forming the surface of the bubble and the new hadron phase. A brief
account of a first order quark-hadron phase transition in the
expanding universe may be envisioned  as follows \cite{kajantie}.
When a hadron bubble is nucleated, latent heat is released and a
spherical shock wave expands into the surrounding supercooled
quark--gluon plasma. The plasma thus formed is  reheated and
approaches the critical temperature, preventing further nucleation
in a region passed by one or more shock fronts. Bubble growth is
generally described by deflagrations where a shock front precedes
the actual transition front. The stopping of the nucleation occurs
when the whole universe has reheated $T_c$. The prompt ending of
this phase transition, in about 0.05 $\mu$s, renders the cosmic
expansion completely negligible over this period. Afterwards, the
hadron bubbles grow at the expense of the quark phase and eventually
percolate or coalesce. Eventually, when all quark-gluon plasma has
been converted into hadrons, neglecting possible quark nugget
production, the transition ends. The physics of the quark–-hadron
phase transition and its cosmological implications have been
extensively discussed in the framework of general relativistic
cosmology in \cite{quark}-\cite{quark12}.

As is well known, the Friedmann equations in brane-world scenarios
differs from that of the standard $4D$ cosmology in that they result
in an increased expansion rate at early times. We expect this
deviation from the standard $4D$ cosmology to have noticeable
effects on the cosmological evolution, especially on cosmological
phase transitions. In the context of brane-world models, the first
order phase transitions have been studied in \cite{DaVe01} where it
has been shown that due to the effects coming from higher
dimensions, a phase transition requires a higher nucleation rate to
complete and baryogenesis and particle abundances could be
suppressed. Recently, the quark-hadron phase transition was studied
in a Randall-Sundrum brane-world scenario \cite{harko}.  Within the
framework of first order phase transitions, the authors studied the
evolution of the relevant cosmological parameters (energy density,
temperature, scale factor, etc.) of the quark-gluon and hadron
phases and the phase transition itself. In another attempt, the
phase transition of quarks and gluons was studied in a brane-world
model in which the confinement of matter fields on the brane is
achieved through a confining potential \cite{heydari}. As was
mentioned above, it would therefore be of interest to study phase
transitions of this nature in the context of a BD brane-world
scenario and this is what we intend to do in what follows. In
addition, as recent calculations in lattice QCD strongly favor a
smooth crossover transition \cite{tan}, the study of the formation
of hadrons in such a scenario seems to be of particular importance
for a better understanding of the subtleties of the early universe
and this is what we shall present in the last section.

\section{Field equations in Brans-Dicke brane scenario}
We start by writing the action for the Brans-Dicke brane-world
\cite{mendes}
\begin{equation}\label{eq2.1}
{\cal S}=-\frac{1}{2\kappa_{_{(5)}}^{2}}\int d^{5}x\sqrt{-g}\left(\phi{\cal R}-\frac{\omega}{\phi}\partial_A\phi\partial^{A}\phi\right)+ \int d^{5}x\sqrt{-g}{\cal L}_{m},
\end{equation}
where ${\cal R}$ is the Ricci scalar associated with the
5-dimensional space-time metric $ g_{_{AB}}$, $\phi$ is a scalar
field which we shall call the BD field, $\omega$ is a dimensionless
coupling constant which determines the coupling between gravity and
the BD scalar field and  ${\cal L}_{\rm m}$ represents the
Lagrangian for the matter fields. Latin indices denote 5-dimensional
components ($A,B=0,\ldots,5$) and for convenience we choose
$\kappa_{(5)}^2=8\pi G_{(5)}=1 $. The variation of the action with
respect to $g_{_{AB}}$ and $\phi$ yields the field equations
\begin{eqnarray}\label{eq2.2}
G_{AB}\equiv{\cal R}_{AB}- \frac{1}{2}  g_{_{AB}} {\cal R} & = &
\frac{1}{\phi}\left[ T_{AB}^{\phi}+T_{AB}\right],
\end{eqnarray}
where
\begin{eqnarray}\label{eq2.3}
T_{AB}^{\phi}=\frac{\omega}{\phi} \left[\phi_{;A}\phi_{; B}
-\frac{1}{2}  g_{_{AB}}\phi_{; C}\phi^{;
C}\right]+\left[\phi_{_{;AB}}- g_{_{AB}} {\phi^{;C}}_{;C} \right]
\,,
\end{eqnarray}
\begin{eqnarray}\label{eq2.4}
\square \phi  & = & \frac{T}{3\omega +4}\,,
\end{eqnarray}
and ${ T= T^{C}}_{C}$ is the trace of the energy-momentum tensor of
the matter content of the $5$-dimensional space-time. Note the
factor $3\omega+4$ in the denominator on the right hand side of the
BD field equation instead of the familiar $2\omega+3$ in the
$4$-dimensional case~\cite{barrow}. This is determined by requiring
the validity of the equivalence principle in our setup, see
\cite{wein} for a discussion of this topic in the context of
$4$-dimensional BD theory.

Being interested in cosmological solutions, we take the spatially
flat cosmology ($k=0$) and consider a $5$-dimensional flat metric of
the following form
\begin{eqnarray}
\label{met1}
ds^2=-n^2(\tau,y)d\tau^2+a^{2}(\tau,y)\delta_{ij}dx^{i}dx^{j} +
b^2(\tau,y) dy^2\,,
\end{eqnarray}
where $i,j =1,2,3$. We also assume an orbifold symmetry along the
fifth direction $y \rightarrow -y$. Next we define the
energy-momentum tensor
\begin{eqnarray}
\label{em0} {T^{\rm A}}_{\rm B}= {T^{\rm A}}_{\rm B}\arrowvert_{\rm
bulk} + {T^{\rm A}}_{\rm B} \arrowvert_{\rm brane}\,,
\end{eqnarray}
where the subscripts ``brane'' and ``bulk'' refer to the
corresponding energy-momentum tensors. For simplicity we assume that
the bulk is devoid of matter other than the BD scalar field. The
brane matter field is held at $y=0$ with the following energy
momentum tensors
\begin{eqnarray}
\label{em1} {T^{\rm A}}_{\rm B}\arrowvert_{\rm brane}=
\frac{\delta(y)}{b}{\rm diag}(-\rho,p,p,p,0)~~~~~~~~\mbox{and}~~~~~~~~
{T^{\rm A}}_{\rm B}\arrowvert_{\rm bulk}={\rm diag}(0,0,0,0,0),
\end{eqnarray}
where $\rho=\rho^{b}+\lambda$ and $p=p^{b}-\lambda$. The above
expressions are written assuming that the brane has ordinary matter
with  tension $\lambda$ and that the bulk is empty. There are
several constraints suggested for the brane tension $\lambda$. One
is that suggested by the big bang nucleosynthesis,  $\lambda\geq1$
MeV$^{4}$ \cite{martens}. A much stronger bound for $\lambda$ is due
to the null results of submillimeter tests of Newton's law, giving
$\lambda\geq10^{8}$ GeV$^{4}$ \cite{martens2}. An astrophysical
lower limit on $\lambda$ which is independent of the Newton's law
and  cosmological limits has been studied in \cite{martens}, leading
to the value $\lambda>5\times10^{8}$ MeV$^{4}$ which is constraint
we will be using in our model.

Using  metric (\ref{met1}) we are now able to write the equations of
motion. The $(0,0)$ component reads
\begin{eqnarray}
\label{eq:00}
3 \left[\frac{\dot a}{a}\left(\frac{\dot
a}{a}+\frac{\dot b}{b}\right)
 - \frac{n^2}{b^2} \left(\frac{a^{\prime
\prime}}{a}+\frac{a^{\prime}}{a}\left(\frac{a^{\prime}}{a}-
\frac{b^{\prime}}{b}\right)\right)\right] &=&
\frac{1}{\phi}\left[T^{\phi}_{\rm 00}+T_{00}\right] \,,
\end{eqnarray}
where
\begin{eqnarray}\label{Teq:00}
 T_{\rm 00}^{\phi}&=&-\dot{\phi}\left(3\frac{\dot{a}}{a} +
\frac{\dot{b}}{b} -\frac{\omega}{2} \frac{\dot{\phi}}{\phi}\right) +
\left( \frac{n}{b} \right)^2 \left[ \phi^{\prime\prime} +
\phi^{\prime} \left( 3 \frac{a^{\prime}}{a} - \frac{b^{\prime}}{b} +
\frac{\omega}{2} \frac{\phi^{\prime}}{\phi} \right)\right].
\end{eqnarray}
The $(i,j)$ components are given by
\begin{center}
\begin{eqnarray}\label{eq:ij}
\left\{-2\frac{\ddot a}{a} -\frac{\ddot b}{b} + \left[\frac{\dot
a}{a} \left(-\frac{\dot a}{a}+2\frac{\dot n}{n} \right)+ \frac{\dot
b}{b} \left(-2\frac{\dot a}{a}+\frac{\dot n}{n}
\right)\right]\right\} \delta_{ij}+\nonumber\\
\left\{\left(\frac{n}{b}\right)^2 \left[2\frac{a^{\prime
\prime}}{a}+\frac{n^{\prime \prime}}{n} + \frac{a^{\prime}}{a}
\left(\frac{a^{\prime}}{a}+2\frac{ n^{\prime}}{n}\right)-
\frac{b^{\prime}}{b} \left(\frac{n^{\prime}}{n}+2\frac{a^{\prime}}
{a}\right) \right]\right\} \delta_{ ij}= \frac{1}{\phi}\left(
\frac{n}{a}\right)^2\left[T^{\phi}_{ij}+T_{ij}\right],
\end{eqnarray}
\end{center}
where
\begin{eqnarray}\label{Teq:ij}
T_{ij}^{\phi}&=&\left\{ \frac{\ddot{\phi}}{\phi} +
\frac{\dot{\phi}}{\phi} \left(2
    \frac{\dot{a}}{a} + \frac{\dot{b}}{b}- \frac{\dot{n}}{n} +
    \frac{\omega}{2} \frac{\dot{\phi}}{\phi} \right) -
  \left(\frac{n}{b}\right)^2 \left[ \frac{\phi^{\prime\prime}}{\phi} +
    \frac{\phi^{\prime}}{\phi} \left(2
    \frac{a^{\prime}}{a}  \frac{b^{\prime}}{b}+ \frac{n^{\prime}}{n} +
    \frac{\omega}{2} \frac{\phi^{\prime}}{\phi} \right)\right]
  \right\} \delta_{ij}
\end{eqnarray}
The $(0,5)$ component takes the form
\begin{eqnarray}
\label{eq:05} 3 \left(\frac{\dot{a}}{a}\frac{n^{\prime}}{n} +
\frac{\dot{b}}{b}
  \frac{a^{\prime}}{a} -
  \frac{\dot{a}^{\prime}}{a}\right) =\frac{1}{\phi}T^{\phi}_{05}\,,
\end{eqnarray}
where
\begin{equation}\label{Teq:ijij}
T_{05}^{\phi}=\dot{\phi}^{\prime} - \dot{\phi}\left(
  \frac{n^{\prime}}{n} - \omega\frac{\phi^{\prime}}{\phi}
\right) - \frac{\dot{b}}{b} \phi^{\prime}.
\end{equation}
Finally, for the $(5,5)$ component one has
\begin{eqnarray}
\label{eq:55} 3 \left[-\left(\frac{\ddot a}{a} + \frac{\dot
a}{a}\left(\frac{\dot a}{a} - \frac{\dot n}{n}\right)\right) +
\left(\frac{n}{b} \right)^2 \left(
\frac{a^{\prime}}{a}\left(\frac{a^{\prime}}{a} +
\frac{n^{\prime}}{n}\right)\right) \right]=\frac{1}{\phi}\left(
\frac{n}{b}\right)^2\left[T^{\phi}_{55}+T_{55}\right],
\end{eqnarray}
where
\begin{equation}\label{Teq:55}
T_{55}^{\phi}=\ddot{\phi} +\dot{\phi} \left( 3 \frac{\dot{a}}{a} -
\frac{\dot{n}}{n} + \frac{\omega}{2}
 \frac{\dot{\phi}}{\phi} \right) - \left( \frac{n}{b}\right)^2
\phi^{\prime} \left( 3\frac{a^{\prime}}{a}  + \frac{n^{\prime}}{n} -
\frac{\omega}{2} \frac{\phi^{\prime}}{\phi} \right).
\end{equation}
The equation of motion for the BD field reads
\begin{eqnarray}
\label{eq:bdfield} \ddot{\phi} + \dot{\phi} \left( 3
\frac{\dot{a}}{a} +
  \frac{\dot{b}}{b} - \frac{\dot{n}}{n}\right) - \left( \frac{n}{b}
\right)^2 \left[ \phi^{\prime\prime} + \phi^{\prime} \left( 3
    \frac{a^{\prime}}{a} - \frac{b^{\prime}}{b} + \frac{n^{\prime}}{n}
  \right)\right] & = & - n^2 \frac{T}{3
  \omega+4},
\end{eqnarray}
where a dot represents the time derivative with respect to $\tau$
and the prime corresponds to derivatives with respect to $y$. Note
that in the above derivation we have assumed $\phi=\phi(\tau,y)$. We
make the assumption that the metric and the BD field are continuous
across the brane localized at $y=0$. However, their derivatives can
be discontinuous at the brane position in the $y$ direction. This
suggests the second derivatives of the scale factor and the BD field
will have a Dirac delta function associated with the positions of
the brane. Since the matter is localized on the brane it will
introduce a delta function in the Einstein equations which will be
matched by the distributional part of the second derivatives of the
scale factor and BD field. For instance at $y = 0$, we have
\cite{mendes}
\begin{eqnarray}\label{eq:jumpa1}
f''=\widehat{f''}+[f']\delta(y),
\end{eqnarray}
where the hat marks the non-distributional part of the
second derivative of the quantity. The part associated
with a delta function, $[f' ]$, is a jump in the derivative
of $f$. Here $f$ could be any of the three quantities $a,$ $n$
or $\phi$. The jump in $f$ at $y = 0$ can be written as
\begin{eqnarray}
[f']=f'(0^{+})-f'(0^{-}),
\end{eqnarray}
and the mean value of the function $f$ at $y =0$ is
defined by
\begin{eqnarray}
\sharp f \sharp=\frac{f'(0^{+})-f'(0^{-})}{2},
\end{eqnarray}

After substituting equation (\ref{eq:jumpa1}) in the Einstein field
equations it is possible to find  the jump conditions for $a$ and
$n$ by matching the Dirac delta functions appearing on the left-hand
side of the Einstein equations to the ones coming from the
energy-momentum tensor, equation (\ref{em0}). For the BD field one
has to use equation (\ref{eq:bdfield}) to evaluate the jump
conditions. We therefore find
\begin{eqnarray}
\label{eq:jumpa}
\frac{[a^{\prime}]_{_{0}}}{a_{_{0}} b_{_{0}}} & = &
-\frac{1}{(3\omega+4)\phi_{_{0}}}\Big[p+(\omega +1) \rho\Big] \,, \\
\label{eq:jumpn} \frac{[n^{\prime}]_{_{0}}}{n_{_{0}} b_{_{0}}}&=&\frac{1}{(3\omega+4)\phi_{_{0}}}
\Big[3(\omega +1)p+
(2\omega +3)\rho\Big]\,,\\
\label{eq:jumpbd} \frac{[\phi^{\prime}]_{_{0}}}{\phi_{_{0}}
b_{_{0}}} & = &
\frac{2}{(3\omega+4)\phi_{_{0}}}\gamma\rho\, ,
\end{eqnarray}
where
\begin{eqnarray}\label{gama}
\gamma=\frac{1}{2}(3w_{m}-1),
\end{eqnarray}
with $w_m=\frac{p}{\rho}$ and the subscript $0$ stands for the brane
at $y=0$. The first two conditions, equations (\ref{eq:jumpa}) and
(\ref{eq:jumpn}), are equivalent to Israel's junction conditions in
general relativity (see \cite{bin} for a discussion of its
application in the context of brane-worlds). It is important to note
that the above jump conditions at $y=0$ depend on the energy density
and  pressure component of the brane and the induced curvature on
the brane. Interestingly, for the radiation dominated phase on the
brane, $\rho=3p$, the jump condition for $\phi$ does not vanish and
is proportional to the energy density and  pressure component of the
induced curvature on the brane.

Using the ($0, 0$) component of the Einstein field equations for the
brane located at $y=0$ and the equations representing the jump
conditions (\ref{eq:jumpa}), (\ref{eq:jumpn}) and (\ref{eq:jumpbd})
one gets the Friedmann equation as follows
\begin{align}\label{Frid}
H^2+ \Upsilon\left(H-\frac{\omega}{6}\Upsilon\right)=
\frac{1}{4(3\omega+4)^2\phi^{2}_0}
\left[
\frac{\omega}{6}(3p-\rho)^2+(2+3\omega+\omega^2)\rho^{2}-\omega p\rho-2p^{2}\right],
\end{align}
where $H=\frac{\dot{a}_{_{_{0}}}}{a_{_{0}}}$ and
$\Upsilon=\frac{\dot{\phi}_{0}}{\phi_0}$. Note that  $\rho$ and $p$
consist of two parts, that is $\rho=\rho^{b}+\lambda$ and
$p=p^{b}-\lambda$, where $\lambda$ is tension on the brane. Using
the ($0,5$) component of the Einstein equation and substituting
equations (\ref{eq:jumpa}) and (\ref{eq:jumpn}) we get the
continuity equation for the matter on the brane
\begin{eqnarray}
  \label{eq:cons}
  \dot \rho +3 \left(\rho + p\right)H
= 0 \,.
\end{eqnarray}

While deriving the above equations we have assumed that, from the
point of view of the brane observer, the extra dimension is static,
that is $b=b_0$. We have also fixed the time in such a way that
$n_0=1$, corresponding to the usual choice of time in conventional
cosmology. Taking the mean value of the BD field equation we obtain
an equation of motion for $\phi$ on the brane
\begin{eqnarray}
\label{eq:bdeq}
\frac{\ddot\phi_{_{0}}}{\phi_{_{0}}}+3H\Upsilon=
\frac{\omega (\rho-3p)^2}{(3\omega +4)^2
\phi_{_{0}}^{2}}\,\,.
\end{eqnarray}
Note that in order to obtain equation (\ref{eq:bdeq}) we also have
to assume that the non-distributional part of $\phi^{\prime\prime}$
vanishes, otherwise, a term involving
$\widehat{\phi^{\prime\prime}}$ will appear in the BD field
equation. As we shall see in the next section it is possible to
obtain cosmologically interesting solutions which verify this
condition. Equations (\ref{Frid}) and (\ref{eq:cons}) are dynamical
equations in our BD brane scenario describing the evolution of the
universe. In the next section we shall examine these equations for
the quark-hadron phase transition in the early universe.

\section{Quark-hadron phase transition}
The quark-hadron phase transition is a notion fundamental to the
study of particle physics, particularly in the context of lattice
gauge theories. However, it is an integral part of any study dealing
with the underlying mechanisms responsible for the evolving universe
at its early stages of formation in which a soup of quarks and
gluons interact and undergo a  phase transition to form hadrons. It
is therefore essential to have a overview of the basic ideas before
attempting to use the results obtained from such a phase transition
and apply them to the study of the evolution of the early universe
within the context of the BD brane-world scenario. In this regard, a
well written and concise review can be found in \cite{harko} and the
interested reader should consult it. Here, it would suffice to
mention the results relevant to our study and leave the details of
the discussion to the said reference.

We start from the equation of state of matter in the quark phase
which can generally be given in the form
\begin{equation}\label{eq3.1}
\rho_{_{q}}^{b}=3a_{_{q}}T^{4}+V(T),~~~~~~~~~~~~p_{_{q}}^{b}=a_{_{q}}T^{4}-V(T),
\end{equation}
where $a_{_{q}} = (\pi^{2}/90)g_{_{q}} $, with $g_{_{q}} = 16 + (21/2)N_F + 14.25
= 51.25$ and $N_F = 2$ with $ V (T )$ being the self-interaction
potential. For $V(T)$ we adopt the expression \cite{quark10}
\begin{equation}\label{eq3.2}
V(T)=B+\gamma_T T^{2}-\alpha_{T}T^{4},
\end{equation}
where $B$ is the bag pressure constant,  $\alpha_T = 7\pi ^{2}/20$
and $\gamma_T = m^{2}_s /4$ with $m_s$ the mass of the strange
quark in the range $m_s \in(60–-200)$ MeV. In the case where the
temperature effects are ignorable, the equation of state in the
quark phase takes the form of the MIT bag model equation of state,
$p_{_{q}}^{b} = (\rho_{_{q}}^{b}- 4B)/3$. Results obtained in low energy hadron
spectroscopy, heavy ion collisions and phenomenological fits of
light hadron properties give $B^{1/4}$ between $100$ and $200$ MeV
\cite{LePa92}.

Once the hadron phase is reached,  one takes the cosmological fluid
with energy density $\rho_{_{h}}^{b}$ and pressure $p_{_{h}}^{b}$ as an ideal gas of
massless pions and nucleons obeying the Maxwell-–Boltzmann
statistics. The equation of state can be approximated by
\begin{equation}\label{eq3.3}
p_{_{h}}^{b}(T)=\frac{1}{3}\rho_{_{h}}^{b}(T)=a_\pi T^4,
\end{equation}
where $a_\pi =(\pi^2/90)g_h$ and $g_h=17.25$. The critical
temperature $T_c$ is defined by the condition $p_{_{q}}(T_c) =
p_{_{h}}(T_c)$ \cite{kajantie}, and is given by
\begin{equation}\label{eq3.4}
T_c=\left[\frac{\gamma_T+\sqrt{\gamma_T^{2}+4B(a_{_{q}}+\alpha_T-a_\pi)}}{2(a_{_{q}}+\alpha_T-a_\pi)}\right]^{1/2}.
\end{equation}
If we take  $m_s = 200$ MeV and $B^{1/4} = 200$ MeV,  the
transition temperature is of the order $T_c \approx 125$ MeV. It
is worth mentioning that since the phase transition is assumed to
be of first order, all the physical quantities exhibit
discontinuities across the critical curve.

\section{Behavior of Brans-Dicke brane universe during quark-hadron phase transition}
We are now in a position to study the phase transition described
above. The framework we are working in is defined by the BD
brane-world scenario for which the basic equations were derived in
section 2. The physical quantities of interest through the
quark-hadron phase transition are the energy density $\rho $,
temperature $T$ and scale factor $a_{_{0}}$. These parameters are
determined by the Friedmann equation (\ref{Frid}), conservation
equation (\ref{eq:cons}) and the equations of state, namely
(\ref{eq3.1}), (\ref{eq3.2}) and (\ref{eq3.3}). To start, we
consider the evolution of the BD brane-world before, during and
after the phase transition era.
\subsection{Behavior of temperature}
Let us consider the era preceding the phase transition for which
$T>T_{c}$ and the universe is in the quark phase. Use of
equations of state of the quark matter and the conservation of
matter on the brane, equation (\ref{eq:cons}),  leads to
\begin{equation}\label{eq4.1}
H=\frac{\dot{a_{_{0}}}}{a_{_{0}}}=-\frac{3a_{q}-\alpha
_{_{T}}}{3a_{q}}\frac{\dot{T}}{T}-\frac{1}{6}\frac{\gamma
_{_{T}}}{a_{q}}\frac{\dot{T}}{T^{3}}.
\end{equation}
Integrating the above equation immediately gives
\begin{equation}\label{eq4.2}
 a_{_{0}}(T)=c~T^{\frac{\alpha _{_{T}}-3a_{q}}{3a_{q}}}\exp \left(
\frac{1}{12}\frac{\gamma _{_{T}}}{a_{q}}\frac{1}{T^{2}}\right) ,
\end{equation}
where $c$ is a constant of integration.

To consider the phase transition in the BD brane model, for
simplicity, we take an {\it ansatz} in the form of a relation
between scale factor on the brane ($a_{_{0}}$) and the BD scalar
field ($\phi_0$) as follows
 \begin{equation}\label{ansatz}
\phi_{_{0}}(\tau)=\mu a_{_{0}}^{n}(\tau),
\end{equation}
where $\mu$ and $n$ are constant. Thus, using  equation
(\ref{ansatz}), the Friedmann equation (\ref{Frid}) can be written
as
\begin{equation}\label{mFrid}
H^2=\frac{1}{2[(3\omega+4)^{^{2}}(2+2n-\frac{\omega}{3}n^2)]\phi_{_{0}}^2}\left[
\frac{\omega}{6}(3p-\rho)^2+(2+3\omega+\omega^2)\rho^{2}-p\rho\omega-2p^{2}\right].
\end{equation}

We may now proceed to obtain an expression describing the evolution
of temperature of the BD brane universe in the quark phase by
combining  equations (\ref{eq3.1}), (\ref{eq3.2}), (\ref{eq4.1}),
(\ref{eq4.2}), (\ref{ansatz}) and (\ref{mFrid}), leading to
\begin{align}\label{eq4.3}
\frac{dT}{d\tau}&=\frac{-T^{(nA_{_{0}}+3)}}{\mu(3\omega+4)(A_{_{0}} T^2+A_{_{1}})
\sqrt{2(2+2n-\frac{\omega}{3}n^2)}}\exp\left(-\frac{nA_{_{_{1}}}}{2T^2}\right)\times\nonumber \\
&\left[
\frac{2\omega}{3}(V(T)-\lambda)^2+(2+3\omega+\omega^2)(\rho_{_{q}}^{b}(T)+\lambda)^{2}-
\omega (p_{_{q}}^{b}(T)-\lambda)(\rho_{_{q}}^{b}(T)+\lambda)-2(p_{_{q}}^{b}(T)-\lambda)^{2}\right]^{1/2}\,,
\end{align}
where we have denoted
\begin{eqnarray}\label{eq4.4}
A_{0}=1-\frac{\alpha_{_{T}}}{3a_{_{q}}},
\\\nonumber
A_{1}=\frac{\gamma_{_{T}}}{6a_{_{q}}}.
\end{eqnarray}
Equation (\ref{eq4.3}) may  be solved numerically and the result is
presented in figure 1 which shows the behavior of temperature as a
function of the cosmic time $\tau$ in a BD brane-world filled with
quark matter for different values of $\omega$ with  $n=0.05$ ,
$\mu=2\times 10^{5}$ and $\lambda=10\times 10^{8}$ MeV$^{4}$.
\begin{figure}
\begin{center}
\epsfig{figure=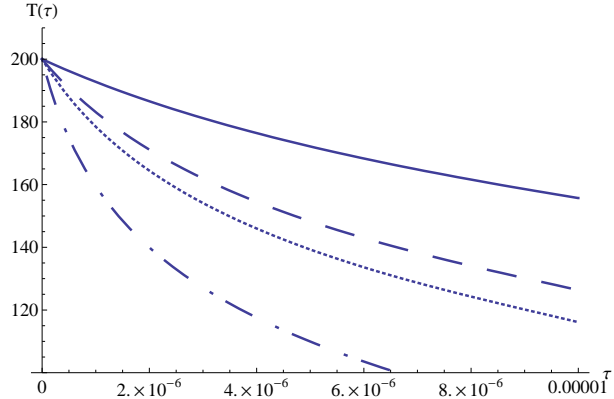,width=8cm}
\end{center}
\caption{\footnotesize  The behavior of $T(\tau)$ as a function of
time ($\tau$) for $\mu=2\times 10^{5}$, $\lambda=10\times 10^{8}$ MeV$^{4}$, $n=0.05$ and different
values of $\omega$: $\omega=1\times 10^{3}$
(solid curve), $\omega=2.3\times 10^{3}$
(dashed curve), $\omega=2.4\times 10^{3}$
(dotted curve) and $\omega=2.5\times 10^{3}$
(dotted-dashed curve). We have taken $B^{1/4}=200$ MeV.}
\end{figure}

\subsection{Temperature behavior with $V(T)=B$}
One may gain considerable insight in the evolution of cosmological
quark matter in our BD brane-world by taking  the simple case in
which temperature corrections can be neglected in the self
interacting potential $V$. Then $V=B={\rm const.}$ and equation of
state of the quark matter is given by that of the bag model, namely
$p_{_{q}}^{b}=\left( \rho _{_{q}}^{b}-4B\right) /3$. Equation
(\ref{eq:cons}) may then be integrated to give the scale factor on
the brane as a function of temperature
\begin{equation}\label{eq4.5}
a_{_{0}}(T)=\frac{c}{T}\,,~~~~~~~~~~~~~~~\phi_{_{0}}=\mu c^{n}T^{^{-n}},
\end{equation}
where $c$ is a constant of integration.

Using equations (\ref{mFrid}) and (\ref{eq4.5}), the time
dependence of temperature can be obtained from  equation
\begin{align}\label{eq:4.6}
\frac{dT}{d\tau}&=\frac{-T^{^{n+1}}}{\mu(3\omega+4)\sqrt{2(2+2n-\frac{\omega}{3}n^2)}}
\times~\nonumber \\
&\left[
\frac{2\omega}{3}(B-\lambda)^2+(2+3\omega+\omega^2)(\rho_{_{q}}^{b}(T)+\lambda)^{2}-\omega\left(\frac{\rho
_{_{q}}^{b}-4B}{3}-\lambda\right)(\rho_{_{q}}^{b}(T)+\lambda)-2\left(\frac{ \rho
_{_{q}}^{b}-4B}{3}-\lambda\right)^{2}\right]^{\frac{1}{2}}\,,
\end{align}
where we have taken $c=1$ and $\rho_{_{q}}(T)$ is given by
\begin{equation}\label{eq4.10}
\rho_{_{q}}^{b}(T)=3a_{_{q}}T^{4}+B.
\end{equation}

The above equation  can be solved numerically and the result is
given in figure 2 which shows the behavior of temperature as a
function of cosmic time $\tau$ in a BD brane-world filled with quark
matter for different values of $\omega$ with $\mu=2\times 10^{5}$ ,
$n=0.05$ and $\lambda=10\times 10^{8}$ MeV$^{4}$ as the
self-interacting potential, $V(T)$, is considered to be a constant.
\begin{figure}
\begin{center}
\epsfig{figure=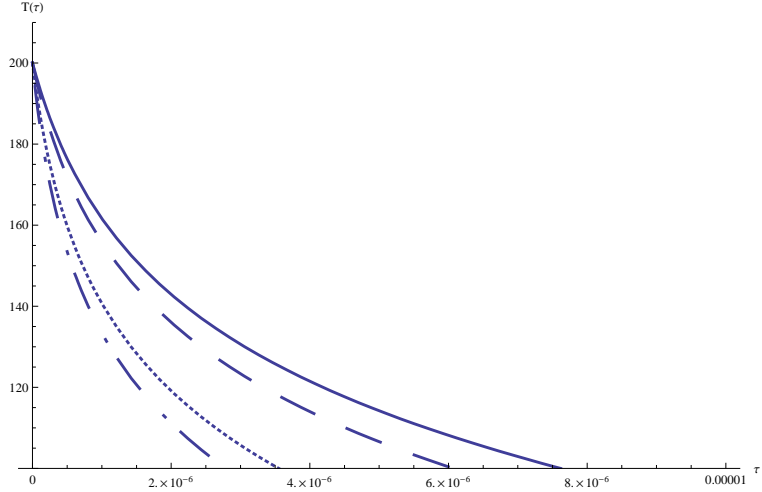,width=10cm}
\end{center}
\caption{\footnotesize  The behavior of $T(\tau)$ as a function of
time ($\tau$) for  $\mu=2\times 10^{5}$, $\lambda=10\times 10^{8}$ MeV$^{4}$, $n=0.05$ and different
values of $\omega$: $\omega=100$
(solid curve), $\omega=1\times 10^{3}$
(dashed curve), $\omega=2\times 10^{3}$
(dotted curve) and $\omega=2.2\times 10^{3}$
(dotted-dashed curve). We have taken $B^{1/4}=200$ MeV.}
\end{figure}
\subsection{Formation of hadrons}
During the phase transition, the temperature and pressure are
constant and quantities like the  entropy $S=sa^{3}$ and enthalpy
$W=\left( \rho +p\right) a^{3}$ are conserved. Also, during the
phase transition $\rho^{b}(t)$ decreases from $\rho _{q}^{b}(T_{c}) \equiv
\rho_{Q}$ to $\rho_{ _{h}}^{b}(T_{c}) \equiv \rho_{_{H}}$ . For phase
transition temperature of $T_{c} = 125$ MeV we have $\rho_{ Q}
\approx 5\times10^{9} $MeV$^4$  and $\rho_{ H} \approx 1.38\times
10^{9} $ MeV$^4$, respectively. For the same value of the
temperature the value of the pressure of the cosmological fluid
during the phase transition is $p_{c}^{b} \approx 4.6 \times 10^{8}$
MeV$^4$. Following \cite{kajantie,harko}, we replace $\rho^{b} \left(
\tau\right)$ by $h(\tau)$, the volume fraction of matter in the
hadron phase, by defining
\begin{equation}\label{eq4.11}
\rho^{b} \left( \tau\right) =\rho _{_{H}}h(\tau)+\rho _{_{Q}}\left[ 1-h(\tau)\right]
=\rho _{_{Q}} \left[ 1+mh(\tau)\right] ,
\end{equation}
where $m=\left( \rho _{_{H}}-\rho _{_{Q}}\right) /\rho _{_{Q}}$. The
beginning of the phase transition is characterized by
$h(\tau_{c})=0$ where $\tau_{c}$ is the time representing it and
$\rho ^{b}\left( \tau_{c}\right) \equiv \rho _{_{Q}}$, while the end
of the transition is characterized by $h\left( \tau_{h}\right) =1$
with $\tau_{h}$ being the time signaling the end and corresponding
to $\rho ^{b}\left( \tau_{h}\right) \equiv \rho _{_{H}}$. For
$\tau>\tau_{h}$ the universe enters into the hadronic phase.

Equation (\ref{eq:cons}) now gives
\begin{equation}\label{eq4.12}
\frac{\dot{a_{_{0}}}}{a_{_{0}}}=-\frac{1}{3}\frac{\left( \rho _{_{H}}-\rho
_{_{Q}}\right) \dot{h} }{\rho _{_{Q}}+p_{c}+\left( \rho _{_{H}}-\rho
_{_{Q}}\right) h}=-\frac{1}{3}\frac{r \dot{h}}{1+rh},
\end{equation}
where we have denoted $r=\left( \rho _{_{H}}-\rho _{_{Q}}\right)
/\left( \rho _{_{Q}}+p_{c}\right)$. The relation between the scale
factor on the brane and the hadronic fraction $h(\tau)$ may now be
obtained from the above equation
\begin{equation}\label{eq4.13}
a_{_{0}}(\tau)=a_{_{0}}\left( \tau_{c}\right) \left[ 1+rh(\tau)\right] ^{-1/3},
~~~~~~~~\phi_{_{0}}=\mu a_{_{0}}^{n}\left( \tau_{c}\right)\left[ 1+rh(\tau)\right] ^{-n/3},
\end{equation}
where use has been made of the initial condition
$h\left(\tau_{c}\right)=0$. Now, using equations (\ref{mFrid}) and
(\ref{eq4.13}) we obtain the time evolution of the matter fraction
in the hadronic phase
\begin{align}\label{eq4.14}
\frac{dh}{d\tau}&=-\frac{3[1+rh(t)]^{\frac{3+n}{3}}}{\mu r(3\omega+4)a_{_{0}}^{n}
\left( \tau_{c}\right)}\frac{1}{\sqrt{2(2+2n-\frac{\omega}{3}n^2)}}
\times~~~\nonumber \\
&\left[\frac{2\omega}{3}(3p_{_{c}}-\chi(t)\rho_{_{_{Q}}}-4\lambda)^{2}+
(2+3\omega+\omega^2)(\chi(t)\rho_{_{Q}}+\lambda)^{2}-\omega (p_{_{c}}-\lambda)(\chi(t)\rho_{_{Q}}+\lambda)-2(p_{_{c}}-\lambda)^{2}\right]^{\frac{1}{2}}\,.
\end{align}
Here we take $a_{_{0}}^{n}( \tau_{c})=1$ and $\chi(t)=1+mh(t)$.
Figure 3 shows variation of the hadron fraction $h(\tau)$ as a
function of $\tau$  for different values of $\omega$ with $\mu=20000$ , $n=0.05$ and $\lambda=8\times 10^{8}$ MeV$^{4}$.
\begin{figure}
\begin{center}
\epsfig{figure=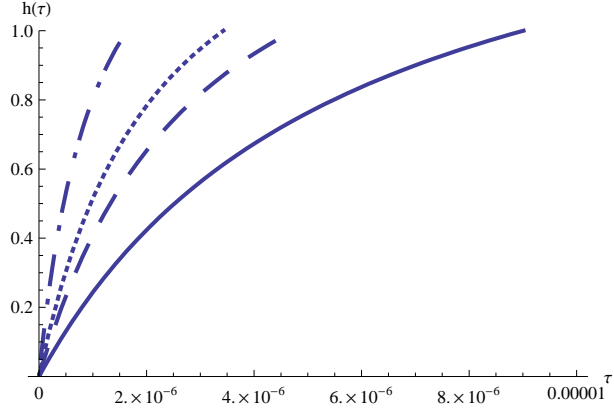,width=8cm}
\end{center}
\caption{\footnotesize  The behavior of $h(\tau)$ as a function of
time ($\tau$) for $\mu=20000$, $\lambda=8\times 10^{8}$ MeV$^{4}$, $n=0.05$, $V(T)=B$ and different
values of $\omega$: $\omega=1\times 10^{3}$
(solid curve), $\omega=2.1\times 10^{3}$
(dashed curve), $\omega=2.3\times 10^{3}$
(dotted curve) and $\omega=2.47\times 10^{3}$
(dotted-dashed curve). We have taken $B^{1/4}=200$ MeV.}
\end{figure}

\subsection{Pure hadronic era}
After the phase transition, the energy density of the pure hadronic
matter is given by $\rho _{h}^{b}=3p_{h}^{b}=3a_{\pi }T^{4}$. The
conservation equation on the brane (\ref {eq:cons}) leads to
\begin{equation}
a_{_{0}}(T)=a_{_{0}}\left( t_{h}\right)
T_{c}/T\,,~~~~~~~~~~~~~\phi_{0}= \mu a^{n}_{_{0}}( t_{h})(
T_{c}/T)^n.
\end{equation}
The temperature dependence of the Brans--Dicke brane universe in the hadronic phase is
governed by the equation
\begin{equation}\label{eq4.19}
\frac{dT}{d\tau}=-\frac{T^{n+1}\left[\frac{8}{3}\omega\lambda^{2}+(\omega^2+\frac{8}{3}\omega+\frac{16}{9})(a_{_{\pi}}T^{4}+\lambda)^{2}\right]^{\frac{1}{2}}}
{\mu [T_{c}a_{_{0}}( t_{h})]^{n}(3\omega+4)\sqrt{2(2+2n-\frac{\omega}{3}n^2)}}.
\end{equation}
Variation of temperature of the hadronic fluid filled BD brane
universe as a function of $\tau$ for different values of $\omega$ with $\mu=20000 $ ,
$n=0.02$ and $\lambda=5\times 10^{8}$ MeV$^{4}$ is represented in figure 4.
\begin{figure}
\begin{center}
\epsfig{figure=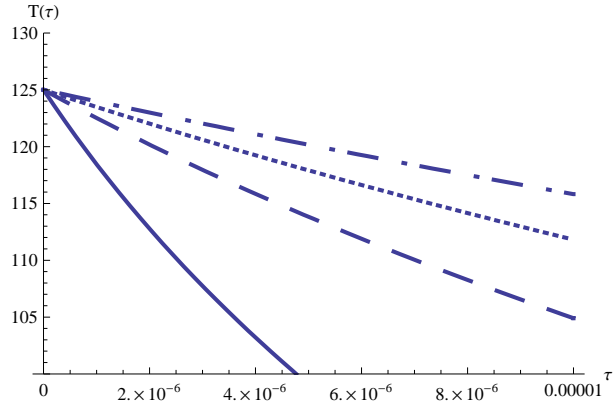,width=8cm}
\end{center}
\caption{\footnotesize  The behavior of $T(\tau)$ as a function of
time ($\tau$) for  $\mu=20000$, $\lambda=5\times 10^{8}$ MeV$^{4}$, $n=0.02$ and different
values of $\omega$: $\omega=15\times 10^{3}$
(solid curve), $\omega=13\times 10^{3}$
(dashed curve), $\omega=9\times 10^{3}$
(dotted curve) and $\omega=1\times 10^{3}$
(dotted-dashed curve). We have taken $B^{1/4}=200$ MeV and $T_{c}=125$ MeV.}
\end{figure}
\subsection{Effects of the BD coupling on phase transition}
It is important to emphasize that the BD coupling, $\omega$, plays
an essential role in the model at hand and it is thus appropriate at
this point to consider the effects of $\omega$ during the phase
transition.

We know that in the early universe the energy density is extremely
high and thus in the high density regime the Hubble function in the
DB brane world is proportional to both the energy density of the
cosmological matter and $\omega$, as can be seen from terms in the
square bracket in equation (\ref{Frid}). Moreover, in the early
universe the BD coupling, which is an indication of the strength of
the DB scalar field, is expected to play an important role and
therefore its numerical value would characterize its influence. In
general, in the limit $\omega\rightarrow\infty$, we recover the FRW
equation in the Randall-Sundrum (RS) model \cite{bin} from equation
(\ref{Frid}).

In the context of the BD brane world model, we have found that the
temperature evolution of the universe is different from that of the
RS brane world model. The temperature of the early universe in the
quark phase is higher in the BD brane world scenario, as can be seen
from the left graph in figure 5, where the dotted-dashed curve, to a
high degree, corresponds to the RS brane world limit of large
$\omega$. Hence a large value of the BD coupling, $\omega$, would
significantly reduce the temperature of the quark–-gluon plasma, and
accelerates the phase transition to the hadronic era. Once the
quark--hadron phase transition starts, the hadron fraction $h$ is
again strongly dependent on the BD coupling $\omega$. From the right
graph in figure 5 it is seen that for large values of $\omega$,
$h(\tau)$ is much higher (dotted-dashed curve) than in the RS model
and standard general relativity. The effect of an increase in
$\omega$ on the brane is to strongly accelerate the formation of the
hadronic phase and shorten the time interval necessary for the
transition. A large $\omega$ tends to reduce the temperature of the
hadronic fluid. From these figures it can be seen that for small
values of $\omega$, the rate of the phase transition is slower than
the large values of $\omega$.
\begin{figure}
\begin{center}
\epsfig{figure=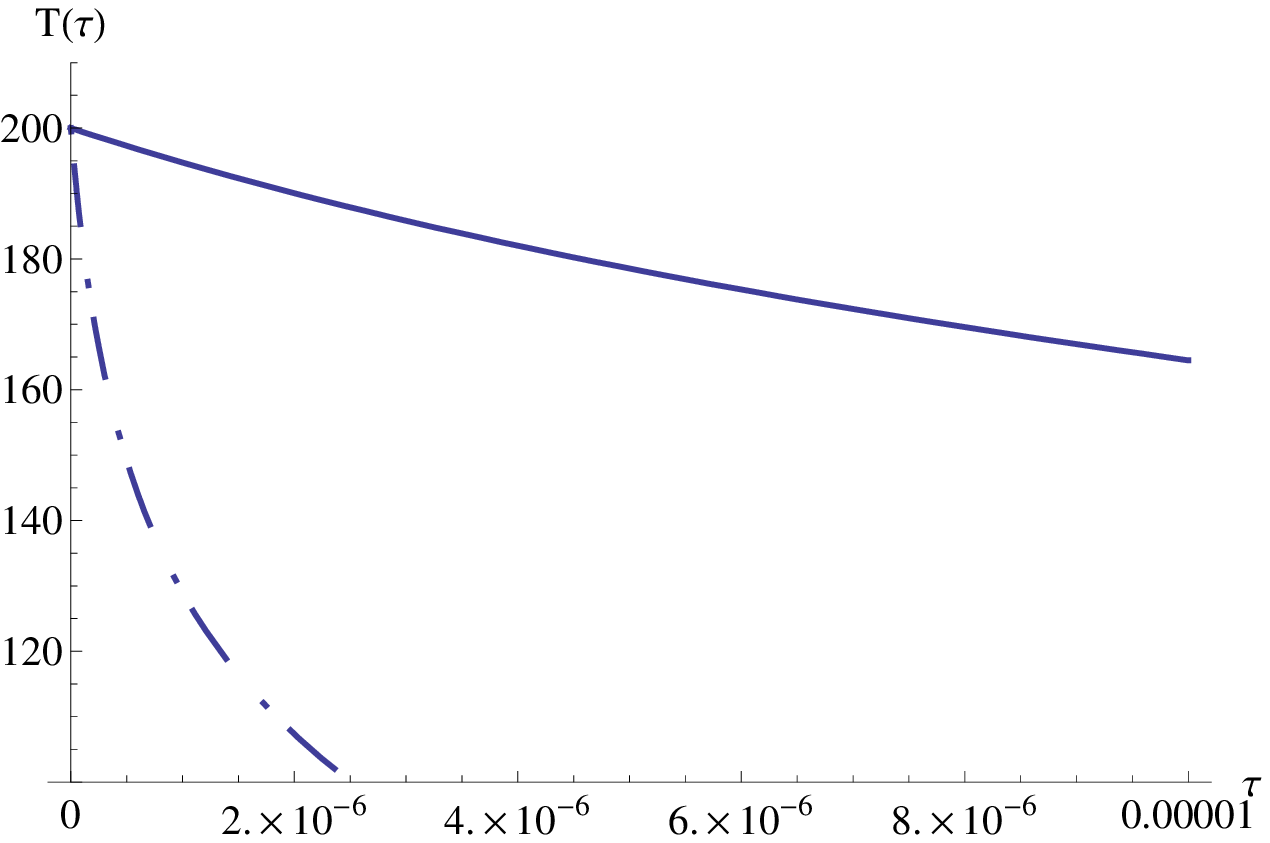,width=7cm}\hspace{5mm}
\epsfig{figure=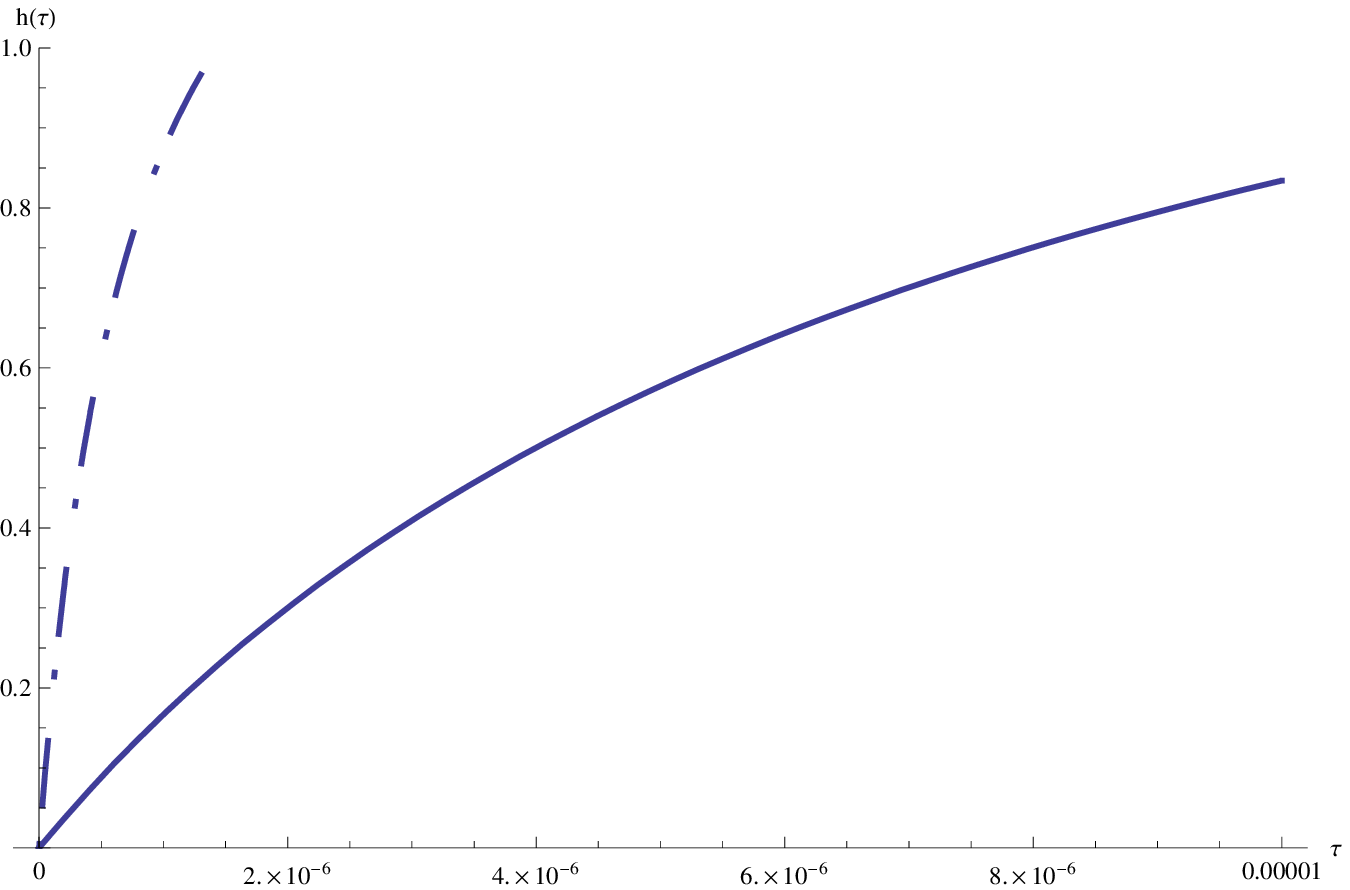,width=7cm}
\end{center}
\caption{\footnotesize  Left, the behavior of $T(\tau)$ as a
function of time ($\tau$) for $\mu=2\times 10^{5}$,
$\lambda=10\times 10^{8}$ MeV$^{4}$, $n=0.05$ and small and large
values of $\omega$, respectively : $\omega=1\times 10^{0}$ (solid
curve) and $\omega=2.518\times 10^{3}$ (dotted-dashed curve). Right,
the behavior of $h(\tau)$ as a function of time ($\tau$) for
$\mu=20000$, $\lambda=8\times 10^{8}$ MeV$^{4}$, $n=0.05$, $V(T)=B$
and small and large values of $\omega$, respectively:
$\omega=1\times 10^{1}$ (solid curve) and $\omega=2.5\times 10^{3}$
(dotted-dashed curve). We have taken $B^{1/4}=200$ MeV .}
\end{figure}
\section{Lattice QCD phase transition}
As was mentioned in the introduction, lattice QCD calculations for
two quark flavors suggest that QCD makes a smooth crossover
transition at a temperature of $T_c \sim 200$ MeV \cite{tan}. It is
therefore necessary to have a brief review of the basic notions of
this subject before using the results to study the universe at early
times within the context of Brans-Dicke brane scenario without the
cosmological constant, $\lambda$, on the brane.

Lattice QCD is an approach which allows one to systematically study
the non-perturbative regime of the QCD equation of state. This
approach has enabled the calculation of the QCD equation of state
using supercomputers \cite{cheng} with two light quarks and a
heavier strange quark on a $(N_t = 6)$ $\times 32^3$ size lattice.
The quark masses have been chosen to be close to their physical
value, i.e. the pion mass is about $220$ MeV . The equation of state
was calculated at a temporal extent of the lattice $N_t = 6$ for
which sizable lattice cut-off effects are still present
\cite{gupta}. The data for energy density $\rho(T)$, pressure $p(T)$
and trace anomaly $\rho- 3p$ and entropy $s$, used in what follows
are taken from reference \cite{cheng}. We also note that besides the
strange quark, one can also include the effect of the charm quark as
well as photons and leptons on the equation of state. These have
important cosmological contributions as was shown in \cite{laine}.
Recent references on lattice QCD at high temperature can be found in
\cite{cheng2}.

Over the high temperature regime,  radiation like behavior is seen
as expected. However, in the region at and below the critical
temperature $T_c$ ($\approx200$ MeV ) of the deconfinement
transition, the behavior changes drastically. This behavior change
is also relevant for cosmological observables as we will see in the
following. For high temperature, that is between 2.82 (100MeV) and
7.19 (100MeV), one can fit the data to a simple equation of state
of the form
\begin{eqnarray}\label{eos}
\rho(T)\approx\alpha T^4,\\\nonumber p(T)\approx\sigma T^4.
\end{eqnarray}
The values of $\alpha = 14.9702\pm009997$ and $\sigma =
4.99115\pm004474$ are found using a least squares fit \cite{cheng}.

While for times before the phase transition the lattice data matches
the radiation behavior very well, for times corresponding to
temperatures above $T_c$ the behavior of the lattice data changes
towards matter dominated behavior. We remark that lattice studies
show that the QCD phase transition at its physical values is
actually a crossover transition.

\subsection{High Temperature Regime}
Let us first consider the era before phase transition at high
temperature  where the universe is in the quark phase. Using the
conservation equation of matter together with the  equation of state
of quark matter (\ref{eos}), one finds the following relation for
the Hubble parameter
\begin{equation}\label{eq15}
H=\frac{\dot{a}}{a}=-\frac{4\alpha}{3(\alpha+\sigma)}\frac{\dot{T}}{T},
\end{equation}
whose solution is given by
\begin{equation}\label{eq16}
a(T)=cT^{\frac{-4\alpha}{3(\alpha+\sigma)}},
\end{equation}
where $c$ is a constant of integration.

One may now proceed to obtain an expression describing the behavior
of temperature of the Brans-Dicke brane universe with respect to
time in the quark phase. Using equations (\ref{mFrid}), (\ref{eos}),
(\ref{eq15}) and (\ref{eq16}) one finds a differential equation for
the temperature
\begin{equation}\label{eq19}
\frac{dT}{d\tau}=-\frac{3(\alpha+\sigma)T}{4\mu\alpha(3\omega+4)
T^{\frac{-4n\alpha}{3(\alpha+\sigma)}}}\left( \frac{\frac{\omega}{6}
(3p - \rho)+(2 + 3 \omega + \omega^2)\rho^2 -\omega p\rho - 2 p^2}{2 [2(n+1)-
\frac{n^{2}\omega}{2}]}\right)^{1/2},
\end{equation}
where we have set the constant $c$ to unity. The transition region
in the crossover regime can be defined as the temperature interval
$282$ MeV $< T < 719$ MeV. In view of the crossover nature of the
finite temperature QCD transition, such definition is equivocal
\cite{pasi}.

Equation (\ref{eq19}) can be solved numerically and the result is
plotted in figure 5 which shows the behavior of temperature of the
Universe in the quark phase as a function of the cosmic time in
Brans-Dicke brane cosmology for $\omega=3\times10^{5}$, in the
interval $282$ MeV $< T < 719$ MeV in the high temperature regime.
We see that as the time evolves the universe becomes cooler.
\begin{figure}[t]
\begin{center}
\epsfig{figure=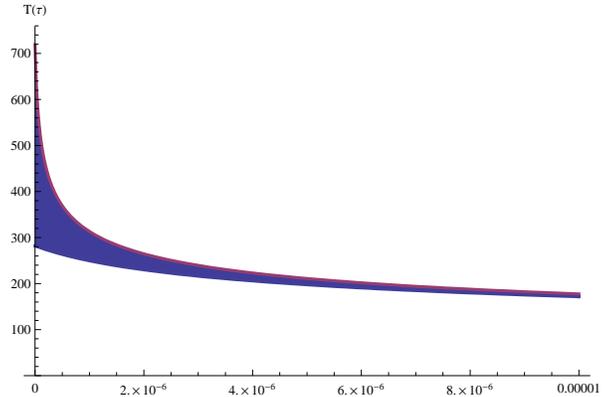,width=8cm}
\end{center}
\caption{\footnotesize The behavior of $T(t)$ in the interval $282$
MeV $< T < 719$ MeV as a function of $\tau$ for $\mu=10^{4}$,
$n=0.00004$ and $\omega=3\times10^{5}$ .}
\end{figure}

\subsection{Low temperature regime}
Besides lattice QCD there are other approaches to the low
temperature equation of state. In the framework of the Hadronic
Resonance Gas model (HRG), QCD in the confinement phase is treated
as a non-interacting gas of fermions and bosons \cite{karsch}. The
fermions and bosons in this model are the hadronic resonances of
QCD, namely mesons and baryons. The idea of the HRG model is to
implicitly account for the strong interaction in the confinement
phase by looking at the hadronic resonances only since these are
basically the relevant degrees of freedom in that phase. The HRG
model is expected to give a good description of thermodynamic
quantities in the transition region from high to low temperature
\cite{p}. The HRG result for the trace anomaly can also be
parameterized by the simple form \cite{pasi}
\begin{equation}\label{HRG}
\frac{\Theta(T)}{T^4}\equiv\frac{\rho-3p}{T{^4}}=a_1T + a_2T^3 + a_3T^4 + a_4T^{10},
\end{equation}
with $a_1 = 4.654$ GeV$^{-1}$, $a_2 = -879$ GeV$^{-3}$, $a_3 = 8081$ GeV$^{-4}$, $a_4 = -7039000$ GeV$^{-10}$.

In lattice QCD the calculation of the pressure, energy density and
entropy density usually proceeds through the calculation of the
trace anomaly $\Theta(T) = \rho(T) - 3p(T)$. Using the thermodynamic
identities, the pressure difference at temperatures $T$ and $T_{\rm
low}$ can be expressed as the integral of the trace anomaly
\begin{equation}\label{trace}
\frac{p(T)}{T^4}-\frac{p(T_{\rm low})}{T^4_{\rm low}}=\int^{T}_{T_{\rm low}}\frac{dT'}{T'^5}\Theta(T').
\end{equation}
By choosing the lower integration limit sufficiently small,
$p(T_{\rm low})$ can be neglected due to the exponential
suppression. Then the energy density $\rho(T) = \Theta(T)+3p(T)$ and
the entropy density $s(T) = (\rho+p)/T$ can be calculated. This
procedure is known as the integral method \cite{boyd}. Using
equations (\ref{HRG}) and (\ref{trace}) we obtain
\begin{eqnarray}\label{rho}
\rho(T)=3\eta T^4 + 4 a_{1}T^5 +2 a_{2}T^7 + \frac{7a_3}{4}T^8 + \frac{13a_4}{10}T^{14},\\\nonumber
p(T)=\eta T^4 +  a_{1}T^5 +\frac{ a_{2}}{3}T^7 + \frac{a_3}{4}T^8 + \frac{a_4}{10}T^{14},
\end{eqnarray}
where $\eta=-0.112$. The trace anomaly plays a central role in
lattice determination of the equation of state. The equation of
state is obtained by integrating the parameterizations given in
equations (\ref{HRG}) over the temperature as shown in equation
(\ref{trace}).

Let us now consider the era before phase transition at low
temperature where the universe is in the confinement phase and is
treated as a non-interacting gas of fermions and bosons
\cite{karsch}. Using the conservation equation of matter together
with  equation of state (\ref{rho}), one gets the following relation
for the Hubble parameter
\begin{equation}\label{eq155}
H=\frac{\dot{a}}{a}=-\frac{12\eta T^3+20a_1 T^4+A(T)}{3[4\eta T^4+5a_1 T^5+B(T)]}\dot{T},
\end{equation}
where
\begin{eqnarray}\label{aa}
A(T)=14a_2 T^6+14a_3 T^7+\frac{91}{5}T^{13},\\\nonumber
B(T)=\frac{7}{3}a_2 T^7+2a_3 T^8+\frac{7}{5}a_4 T^{14}.
\end{eqnarray}
One can solve for the scale factor as
\begin{equation}\label{eq166}
a(T)=\frac{c}{T(75 a_1 T + 35 a_2 T^3 + 30 a_3 T^4 + 21 T^{10} +60 \eta)^{1/3}},
\end{equation}
where c is a constant of integration.

We can obtain an expression describing the behavior of temperature
of the Brans-Dicke brane universe with respect to time in the quark
phase. Upon using equations (\ref{mFrid}), (\ref{rho}),
(\ref{eq155}) and (\ref{eq166}) one finds a differential equation
for the temperature as follows
\begin{equation}\label{eq199}
\frac{dT}{d\tau}=-\frac{3 (4\eta T^4 + 5a_1 T^5 + B)}{\mu(3\omega+4)a^{n}(T)(
12\eta T^3 + 20a_1 T^4 +A)}\left( \frac{\frac{\omega}{6}(3p - \rho)+(2 + 3 \omega + \omega^2)\rho^2 -\omega p\rho - 2 p^2}{2 [2(n+1)-\frac{n^{2}\omega}{2}]}\right)^{1/2},
\end{equation}
where we have set the constant $c$ to unity.

\begin{figure}[t]
\begin{center}
\epsfig{figure=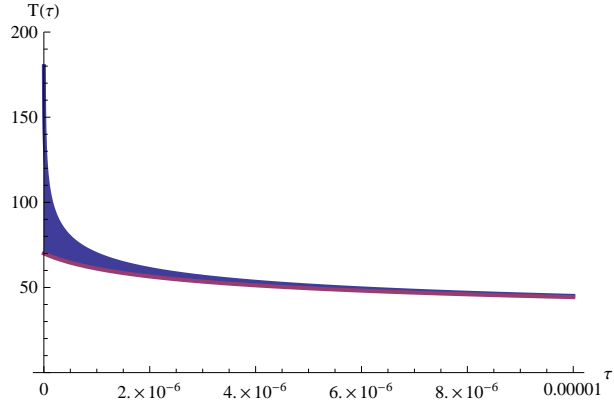,width=8cm}
\end{center}
\caption{\footnotesize The behavior of $T(t)$ in the interval $70$
MeV $< T < 180$ MeV as a function of $\tau$ for $\mu=10^{1}$,
$n=0.00004$, and $\omega=2\times10^{4}$.}
\end{figure}

We can solve  equation (\ref{eq199}) numerically and the result is
plotted in figure  6 which shows the behavior of temperature of the
universe in the quark phase as a function of cosmic time in
Brans-Dicke brane cosmology for $\omega=2\times10^{4}$, in the
interval $70$ MeV $< T < 180$ MeV in the low temperature regime.

The behavior of temperature with time can now be compared in the two
regimes considered above, namely the first order phase transition
and the smooth crossover transition. Figures 2 and 4 show the
variation of temperature in the former while figurers 5 and 6 show
such a variation in the latter. As can be seen, the general behavior
is similar, that is, in both regimes the temperature drops as the
time passes.  We also note that the variation of $T$ with cosmic
time is somewhat different in the two approaches when considered in
detail. In the smooth crossover regime where lattice QCD is used to
investigate the high temperature behavior, we see that the slope is
smooth relative to the first order phase transition while at lower
temperatures where HRG is used, the slope is steep compared to first
order phase transition. Taking into account the energy range in
which the calculations are done, one might conclude that these two
approaches to the quark-hadron transition in the early universe do
not predict fundamentally different ways of the evolution of the
early universe.

\section{Conclusions}
In this paper, we have discussed the quark-–hadron phase transition
in a BD brane cosmological setting in which our universe is a
three-brane embedded in a $5$-dimensional bulk space-time within an
effective model of QCD. We studied  the evolution of the physical
quantities relevant to the physical description of the early
universe; the energy density, temperature and scale factor, before,
during, and after the phase transition. We found that for different
values of $\omega$ phase transition occurs and results in decreasing
the effective temperature of the quark-gluon plasma and of the
hadronic fluid. We then compared our results with the results
presented in \cite{harko} and \cite{heydari}. In \cite{harko} the
authors studied quark--hadron phase transition in a Randall--Sundrum
brane model and showed that for different values of the barne
tension $\lambda$, phase transition occurs. Also in \cite{heydari},
the authors investigated the quark–-hadron phase transition in a
brane-world scenario where the localization of matter on the brane
is achieved through the action of a confining potential and showed
that for different values of the parameters in their model, phase
transition takes place. What has been lacking is a study of such
phase transitions in the context of a  BD brane world model,
presented in this work,  emphasizing the evolution of the physical
quantities relevant to the physical description of the early
universe.

Finally, in section 5 we have considered the quark--hadron
transition in the context of a smooth crossover regime at high and
low temperatures.  Such a study is of particular interest since
extensive studies in lattice QCD over the past few years have led to
the consensus that leans firmly in favor of a smooth crossover
quark--hadron transition in the early universe. We showed that such a
quark-hadron transition occurs and results in decreasing the
effective temperature of the quark-gluon plasma during the process
of the quark--hadron transition. The generic behavior of the
temperature of the early universe in such a scenario is similar to
that of a first order phase transition, although the differences in
the energy should be taken into account.

\section*{Acknowledgment}
K. Atazadeh and H. R. Sepangi would like to thank the research
council of Shahid-Beheshti University for financial support. The
work of A. M. Ghezelbash is supported by the Natural Sciences and
Engineering Council of Canada (NSERC).


\begin{thebibliography}{99}
\bibitem{Rubakov}V. A. Rubakov and M. E. Shaposhnikov, Phys. Lett. B {\bf 125}, (1983) 136;
\\K. Akama, "Pregeometry" in Lecture Notes in Physics, 176, {\it Gauge Theory and Gravitation,Proceedings, Nara,
  1982, (Springer-Verlag)}, edited by K. Kikkawa, N. Nakanishi and H. Nariai, 267;
\\Keiichi Akama, Lect. Notes Phys. {\bf176} (1982) 267, hep-th/0001113.
\bibitem{lisa}L. Randall and R. Sundrum, Phys. Rev. Lett. {\bf 83},(1999) 3370.
\bibitem{mennim}A. Mennim and R. Battye, Class. Quant. Grav. {\bf18} (2001) 2171, hep-th/0008192.
\bibitem{harko} G. De Risi, T. Harko, F. S. N. Lobo and C. S. J. Pun, Nucl. Phys. B {\bf 805} (2008) 190, gr-qc/ 0807.3066.
\bibitem{heydari}M. Heydari-Fard and H. R. Sepangi, Class. Quantum
Grav. {\bf26} (2009) 235021.
\bibitem{Gorenstein}M. I. Gorenstein, V. K. Petrov  and G. M. Zinovjev, Phys. Lett. B {\bf 106} (1981) 327;\\
M. I. Gorenstein , V. K. Petrov, V. P. Shelest and G. M. Zinovev, Theor. Math. Phys. {\bf 52} (1982) 843.
\bibitem{Greiner}M. I. Gorenstein, W. Greiner and Yang Shin Nan, J. Phys. G: Nucl. Part. Phys. {\bf 24} (1998) 725;\\
M. I. Gorenstein, M. Gazdzicki  and W. Greiner,  Phys. Rev. C {\bf72} (1998) w024909;\\
I. Zakout, C. Greiner and J. Schaffner-Bielich, Nucl. Phys. A {\bf781} (2007) 150;\\
I. Zakout and C. Greiner, Phys. Rev. C {\bf78} (2008) 034916;\\
K. A. Bugaev,  Phys. Rev. C {\bf 76} (2007) 014903;\\
K. A. Bugaev, V. K. Petrov and G. M. Zinovjev, arXiv: 0904.4420;\\
A. Bessa, E. S. Fraga  and B. W. Mintz, Phys. Rev. D {\bf79} (2009) 034012.

\bibitem{tan}Y. Aoki, Sz. Borsanyi, S. Durr, Z. Fodor, S. D. Katz, S. Krieg and K. K. Szabo, JHEP {\bf0906} (2009) 088, arXiv: 0903.4155 ;\\
A. Bazavov, {\it et al}, Phys. Rev. D {\bf80} (2009) 014504, arXiv: 0903.4379;\\
L. Ferroni and V. Koch, Phys. Rev. C {\bf79} (2009) 034905;\\
C. De Tar, PoS LATTICE {\bf2008} (2008) 001, arXiv: 0811.2429;\\
Y. Aoki, G. Endrodi, Z. Fodor, S. D. Katz  and K. K. Szabo, Nature {\bf443} (2006) 675;\\
P. de Forcrand and O. Philipsen, JHEP {\bf070} (2007) 077, hep-lat/0607017;\\
Z. G. Tan and A. Bonasera, Nucl. Phys. A {\bf 784}, 368 (2007).
\bibitem{kajantie} K. Kajantie, H. Kurki-Suonio, Phys. Rev. D {\bf 34} (1986) 1719.
\bibitem{quark}J. Ignatius, K. Kajantie, H. Kurki-Suonio and M.
Laine, Phys. Rev.D {\bf 49}(1994) 3854.

\bibitem{quark2}J. Ignatius, K. Kajantie, H. Kurki-Suonio and M.
Laine, Phys. Rev. D {\bf 50} (1994) 3738.

\bibitem{quark3}H. Kurki-Suonio and M. Laine,  Phys. Rev. D {\bf
51}, 5431 (1995).

\bibitem{quark32}H. Kurki-Suonio and M. Laine, Phys. Rev. D {\bf 54}(1996)
7163.

\bibitem{quark33}M. B. Christiansen and J. Madsen, Phys. Rev. D {\bf53}
(1996) 5446.

\bibitem{quark4}L. Rezzolla, J. C. Miller and O. Pantano,  Phys.
Rev. D {\bf 52} (1995) 3202.

\bibitem{quark5}L. Rezzolla and J. C. Miller, Phys. Rev. D {\bf 53}(1996)
5411.

\bibitem{quark6}L. Rezzolla, Phys. Rev. D {\bf 54} (1996) 1345.

\bibitem{quark7}L. Rezzolla, Phys. Rev. D {\bf 54} (1996) 6072.

\bibitem{quark8}A. Bhattacharyya, J. -e. Alam, S. S. P. Roy, B. Sinha, S.
Raha and P. Bhattacharjee, Phys. Rev.  D {\bf61} (2000) 083509.

\bibitem{quark9}A. C. Davis and M. Lilley, Phys. Rev. D {\bf 61} (2000) 043502.
\bibitem{quark10}N. Borghini, W. N. Cottingham and R. Vinh Mau,  J.
Phys. G {\bf 26} (2000) 771.

\bibitem{quark11}H. I. Kim, B.-H. Lee and C. H. Lee, Phys. Rev. D {\bf
64} (2001) 067301.

\bibitem{quark12}J. Ignatius and D. J. Schwarz, Phys. Rev. Lett. {\bf 86} (2001) 2216.

\bibitem{DaVe01} S. C. Davis, W. B. Perkins, A. C. Davis and I. R. Vernon,
 Phys. Rev. D {\bf 63} (2001) 083518.
\bibitem{mendes}L. E. Mendes and A. Mazumdar, Phys. Lett. B {\bf501} (2001) 249.
\bibitem{barrow}J. D. Barrow and J. P. Mimoso, Phys. Rev. D {\bf 50} (1994) 3746.

\bibitem{wein}S. Weinberg,``{\it Gravitation and cosmology: principles and applications of the general theory of relativity},
'' John Wiley \& Sons, 1972.
\bibitem{martens}C. Germani and R. Maartens, Phys. Rev. D {\bf54} (2001) 124010.
\bibitem{martens2}R. Maartens, D. Wands, B. A. Bassett and I. P. C. Heard, Phys. Rev. D {\bf 62} (2000) 041301(R).
\bibitem{bin}P. Binetruy, C. Deffayet and D. Langlois, Nucl. Phys. B {\bf 565} (2000) 269;\\
 P. Binetruy, C. Deffayet, U. Ellwanger and D. Langlois, Phys. Lett. B {\bf 477} (2000) 285.
\bibitem{LePa92}  T. D. Lee and Y. Pang,  Phys. Rept. {\bf 221} (1992)
251.
\bibitem{cheng}M. Cheng et al., Phys. Rev. D {\bf77} (2008) 014511,  arXiv:0710.0354;
\\ M. McGuigan and W. Solner, arXiv:0810.0265.
\bibitem{gupta}R. Gupta,  PoS LAT {\bf2008} (2008) 170.
\bibitem{laine}M. Laine and Y. Schroder,
Phys. Rev. D {\bf73} (2006) 085009, hep-ph/0603048.
\bibitem{cheng2}M. Cheng [RBC-Bielefeld Collaboration], PoS LAT {\bf2007} (2007) 173,  arXiv:0710.4357;\\
G. Endrodi, Z. Fodor, S. D. Katz and K. K. Szabo,  PoS LAT {\bf 2007}(2007) 228,  arXiv:0710.4197;
\\D. E. Miller, Phys. Rept. {\bf443} (2007) 55, hep-ph/0608234.
\bibitem{pasi}P. Huovinen and P. Petreczky, Nucl. Phys. A {\bf837} (2010) 26,  arXiv:0912.2541
\bibitem{karsch}F. Karsch, K. Redlich and A. Tawfik, Eur. Phys. J. C {\bf29} (2003) 549, hep-ph/
0303108;\\
 F. Karsch, K. Redlich and A. Tawfik,  Phys. Lett. B {\bf571} (2003) 67, hep-ph/0306208;\\
 K. Sakthi Murugesan, G. Janhavi and P. R. Subramanian,  Phys. Rev. D {\bf41} (1990) 2384;\\
 A. Tawfik,  Phys. Rev. D {\bf71} (2005) 054502, hep-ph/
0412336.
\bibitem{p}P. Braun-Munzinger, K. Redlich and J. Stachel, nucl-th/0304013;\\A. Andronic, P. Braun-Munzinger and
J. Stachel, Nucl. Phys. A {\bf772} (2006) 167, nucl-th/0511071.
\bibitem{boyd}G. Boyd, J. Engels, F. Karsch, E. Laermann, C. Legeland, M. Lutgemeier and B. Petersson,
Nucl. Phys. B {\bf469} (1996) 419, hep-lat/9602007.
\end{thebibliography}
\end{document}